\documentclass[prb,twocolumn,noshowpacs,amsmath,twoside]{revtex4}
\usepackage{bm}
\usepackage{epsfig}

\begin{document}

\title{Shot-noise spectroscopy of energy-resolved ballistic currents}

\author{M. Naspreda}
\author{O. M. Bulashenko}
\author{J. M. Rub\'{\i}}
\affiliation{Departament de F\'{\i}sica Fonamental,
Universitat de Barcelona, Diagonal 647, E-08028 Barcelona, Spain}
\date{July 17, 2003}

\begin{abstract}
We investigate the shot noise of nonequilibrium carriers injected 
into a ballistic conductor and interacting via long-range Coulomb forces. 
Coulomb interactions are shown to act as an energy analyzer of the profile 
of injected electrons by means of the fluctuations of the potential barrier 
at the emitter contact. We show that the details in the energy profile can be 
extracted from shot-noise measurements in the Coulomb interaction regime, 
but cannot be obtained from time-averaged quantities or shot-noise measurements
in the absence of interactions.
\end{abstract}


\maketitle

\section{Introduction}

The phenomenon of shot noise, associated with the randomness and discreteness
in charge transmission has become a fundamental issue in the study 
of mesoscopic systems in the nanoscale region.
\cite{landauer98,blanter00,vanruitenbeek01}
Since shot noise contains temporal information on the transmitted carriers,
it can be used to deduce, for example, an effective quasiparticle charge, 
kinetic parameters, or other parameters on the interaction 
among carriers---information that usually cannot be obtained 
from time-averaged measurements of the mean current or conductance.
In particular, shot noise is currently used as a tool to probe
fractional charge, \cite{frac} effective superconducting charge, \cite{superc}
quantum transmission modes in atomic-size contacts, \cite{atomcont}
mechanisms of tunneling, \cite{tun} etc. (see also recent review in Ref.
\onlinecite{blanter00}).

A matter of particular interest is the significance of Coulomb interactions
in scattering-free or ballistic conductors. 
As was recently shown, Coulomb interactions may suppress the shot noise down 
to several orders of magnitude in space-charge-limited ballistic 
conductors. \cite{prb00,prb02} 
In this paper, we focus mainly on the question of how the effect of Coulomb 
interactions on the shot noise can be employed to reveal important 
information on the energy profile of nonequilibrium carriers 
injected from an emitter contact.

Using ballistic electrons to study nanoscale structures has recently been
a very active research area.
In the standard technique called ``hot-electron spectroscopy,''
\cite{heiblum85,xie99}
carriers injected from an emitter contact are analyzed in a collector
contact by means of a barrier that is transparent only for carriers
having energy greater than the barrier height.
By changing the bias on the collector barrier, the electron energy profile
can be analyzed. This technique requires the design of a special collector
filter for obtaining information on the electron energies.
Here, we discuss an alternative method that does not require a design of
the filter, rather it employs a ``natural'' filter: the potential barrier
that appears due to an injected space charge.
This space charge limits the current producing the resistance effect by means
of a barrier, which reflects a part of the injected carriers back
to the emitter.
The height of the barrier depends on the screening parameter of the material,
and it varies with the external bias.
The essential difference with the case of a fixed barrier is that
the space-charge barrier fluctuates in time and produces long-range
Coulomb correlations between the transmitted electrons
that leads to the significant suppression of shot noise registered
at the collector contact.
The level of suppression depends drastically on the energy profile
of the injected carriers, \cite{prb00b} while the time-averaged quantities
(the mean current, conductance, etc.) do not.
Therefore, one can use the shot-noise measurements to reveal the details in
the energy profile.

The paper is organized as follows.
In Sec.~II we introduce the basic equations describing 
the space-charge-limited ballistic transport and noise in a two-terminal
conductor.
As a particular example, we address the situation when in addition 
to the Fermi-Dirac injection from each lead, there is an extra 
injection of monoenergetic electrons represented
by a $\delta$ peak in energy spectrum at the emitter lead. 
The self-consistent steady-state solutions for this case are found in Sec.~III.
The formulas for the mean current and noise are obtained in Sec.~IV.
Section V shows the results of the calculations for a GaAs ballistic structure.
Finally, Sec.~VI summarizes the main contributions of the paper.
\begin{figure}[b]
\epsfxsize=8.0cm
\epsfbox{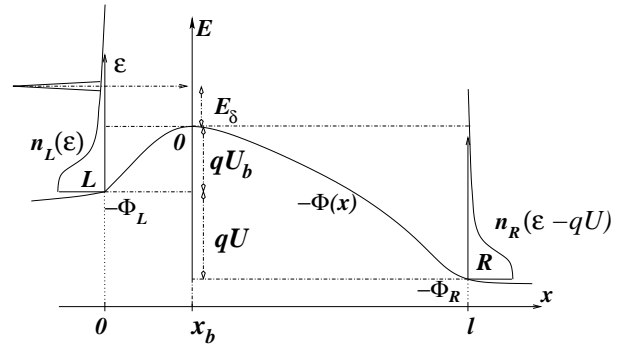}
\caption{Energy diagram determining the potential barrier shape for a ballistic
two-terminal conductor under applied bias $U$. Electrons with energies $E>0$ 
pass over the barrier, while those with $E<0$ are reflected back to the leads.
At the left lead, electrons additional to the Fermi-Dirac distribution are 
injected at energy $E_\delta$ above the barrier.}
\label{fig} \end{figure}

\section{Basic equations}

We consider a two-terminal multimode ballistic conductor in a planar lead
geometry (Fig.~\ref{fig}). 
In a semiclassical framework, the electron occupation numbers $f(x,\vec{k},t)$
are determined by the electron flows from the left and right leads. 
\cite{prb02}
The inhomogeneity of the space charge disturbs the electrostatic potential
in such a way that the self-consistent built-in field determines the potential
barrier, at which electrons are either reflected or transmitted depending on
their energy (Fig.~\ref{fig}).
Assuming that the barrier is much wider than the wavelength 
of electrons, one can neglect tunneling and quantum reflection, i.e.,
the transmission probability is 1 if the electron energy
is higher than the barrier height, and it is 0 in the opposite case.
In this framework, the transport is described by the collisionless Boltzmann 
equation self-consistently coupled with the Poisson equation supplemented by 
the stochastic boundary conditions for the occupation numbers: 
\cite{prb00,prb02}
\begin{gather}\label{boltz}
\left( \frac{\partial}{\partial t} + \frac{\hbar k_x}{m} \frac{\partial}{
\partial x} + q \frac{d \varphi}{dx} \frac{\partial}{\hbar \partial 
k_x} \right) f(x, \vec{k}, t) = 0, \\
\frac{d^2\varphi}{dx^2} = \frac{q}{\epsilon}\int
\frac{d\vec{k}}{(2\pi)^d} f(x, \vec{k}, t), \\
f(0, \vec{k}, t)\big|_{k_x > 0} = f_L(\vec{k}) + \delta f_L
(\vec{k},t), \\
f(\ell,\vec{k}, t)\big|_{k_x<0} = f_R(\vec{k}) + \delta f_R
(\vec{k},t), \\ 
\varphi(\ell, t) - \varphi(0,t) = U(t),
\end{gather}
where $m$ is the electron effective mass, $q$ is the electronic charge, 
$\vec{k} = (k_x, \vec{k}_\perp)$, $d$ is the dimension of a momentum space 
(the spin components are neglected), $\epsilon$ is the dielectric permittivity 
of the media, $\varphi(x,t)$ is the self-consistent electric potential, 
and $U$ is the applied bias (below we assume that $U$ is fixed by an external
circuit). 
In the absence of scattering, the noise originates from the stochastic 
sources $\delta f_{L,R}$ at the left ($L$) and right ($R$) leads,
which gives rise to the fluctuations of the occupation numbers and
electric potential along the conductor and, as a consequence, 
the fluctuations of the current.

Assuming that the number of the occupied transversal modes is large, one can 
integrate the occupation numbers $f(x,\vec{k},t)$ over the transverse momentum
$\vec{k}_\perp$ and obtain for each longitudinal energy $\varepsilon$ 
the (fluctuating) occupation factor at a cross section $x$:
\begin{equation*}  \label{nmod}
n(x,\varepsilon,t) = \int_0^{\infty}
f(x,\varepsilon,\varepsilon_{\perp},t)\
\nu_{\perp}\ d\varepsilon_{\perp},
\end{equation*}
where $\varepsilon$=$\hbar^2 k_x^2/(2m)$, $\varepsilon_{\perp}$=
$\hbar^2 \vec{k}_{\perp}^2/(2m)$, and $\nu_{\perp}=m/2\pi\hbar^2$ 
is the density of transverse modes.

It is seen that under the ballistic transport conditions, the occupation 
factors in the bulk are determined by the occupation factors 
$n_{L,R}(\varepsilon,t)$ in the leads. 
Our aim is to describe how the details of the injection energy profile 
can be revealed in the shot noise. 
To this end, we consider the situation when in addition to the equilibrium 
Fermi-Dirac (FD) injection from each lead (which we call the background 
injection), there is an extra injection from the left (emitter) lead with 
a monoenergetic distribution described by a Dirac $\delta$ function 
(Fig.~\ref{fig}). 
The origin of this additional injection may be related, e.g., to tunneling
from a resonant level (double barrier resonant tunneling emitter), a narrow 
miniband (superlattice emitter), \cite{tun-emit} and cold cathodes.
\cite{cahay}
We also assume that the overall injection current is mostly carried by
the background electrons, while the contribution from the $\delta$ peak 
electrons is relatively small.
Thus for the time-averaged occupation factors in the leads, we write
\begin{align} \label{occup_L}
&n_L(\varepsilon) = n_{FD}(\varepsilon) 
+ a n_0\,k_B T\,\delta(\varepsilon - \varepsilon_\delta), \\
&n_R(\varepsilon) = n_{FD}(\varepsilon-qU), \label{occup_R}
\intertext{where}
&n_{FD}(\varepsilon) = n_0 \ln 
\{1 + \exp [(\varepsilon_F-\varepsilon)/(k_B T)] \}, \label{occup_FD}
\end{align}
is the FD occupation number integrated over the transverse modes,
$n_0={\cal N}/(\xi A)$, ${\cal N}=(k_F^2 A/4\pi)$ is the number of transverse 
modes in the degenerate limit, $\xi=\varepsilon_F/k_BT$,
$\varepsilon_F$ is the Fermi energy at the left lead, $T$ is the temperature,
$A$ is the cross-sectional area, $\varepsilon_\delta$ is the longitudinal 
energy of the extra electrons, and $a$ is the dimensionless peak 
``amplitude.''

Concerning the boundary conditions for the fluctuations of the occupation 
numbers, we assume that the peak electrons are Poissonian, i.e., they
are not correlated. The background FD electrons are correlated 
among themselves according to the Pauli exclusion for the Fermi statistics.
Thus for the energy-resolved injection current fluctuations 
we have \cite{prb02}
\begin{equation*} \label{kuncor}
\langle\delta I_k(\varepsilon)\delta I_k(\varepsilon')\rangle = 
K_k(\varepsilon)(\Delta f)\delta(\varepsilon-\varepsilon'),
\end{equation*}
where $k=L,R$; $\Delta f$ is the frequency bandwidth 
(we assume the low-frequency limit), and
\begin{align} \label{K_bound}
&K_L(\varepsilon) = K_{FD}(\varepsilon) 
+ a K_0\,k_B T\,\delta(\varepsilon - \varepsilon_\delta), \\
&K_R(\varepsilon) = K_{FD}(\varepsilon-qU),
\intertext{with}
&K_{FD}(\varepsilon) = K_0 \,
\{1+\exp [(\varepsilon-\varepsilon_F)/(k_B T)]\}^{-1}, \label{K_FD}
\end{align}
$K_0=2G_S/\xi$, $G_S=G_0{\cal N}$ is the Sharvin conductance, and
$G_0=q^2/(2\pi\hbar)$ is the unit of conductance. \cite{prb02}
The $\delta$ functions in Eqs.~(\ref{occup_L}) and (\ref{K_bound}) implie that
the width of the peak is narrow on the scale of the temperature $T$.
The addition of extra peak electrons bring the injection away from equilibrium.
Therefore one cannot use the thermal-equilibrium Nyquist relationship 
\cite{prb02} for the injected electrons.

\section{Self-consistent steady-state spatial profiles}

It is convenient to introduce the mean total longitudinal energy 
$E=\varepsilon-\Phi(x)$, where the potential energy is counted off from 
the barrier top: $\Phi(x)\equiv q\varphi(x)-q\varphi(x_b)$. 
Then at the leads we obtain $\Phi_L\equiv\Phi(0)=qU_b$ and
$\Phi_R\equiv\Phi(\ell)=\Phi_L+qU$, where $qU_b$ is the barrier height
(Fig.~\ref{fig}).
The solution of Eq.~(\ref{boltz}) for the stationary case gives, after 
integration over the energy, the electron density at any section 
of the conductor in terms of the potential $\Phi$. 
For the boundary conditions (\ref{occup_L})--(\ref{occup_FD}), we find
two additive contributions to the electron density--from the background FD 
electrons and from the $\delta$ peak: $N = N_{bg} + N_{\delta}$.
The former is given by 
\begin{widetext}
\begin{equation} \label{den}
N_{bg}(\Phi) = \int_0^\infty [ n_{FD}(E + \Phi_L) + n_{FD}(E + \Phi_R)] 
\, \nu(E + \Phi) \, dE + 2 \int_{-\Phi}^0 
[\theta_{-\chi} n_{FD}(E + \Phi_L) + \theta_{\chi} n_{FD}(E + \Phi_R) ]
\, \nu(E + \Phi) \, dE,
\end{equation}
\end{widetext}
where $\nu(E)=1/[2\pi\hbar v(E)]$ is the density of states 
with $v=\sqrt{2E/m}$, and $\theta_{\chi}\equiv \theta(\chi)$ 
is the Heaviside function with $\chi=x-x_b$.
The first integral in Eq.~(\ref{den}) corresponds to the electrons
transmitted over the barrier ($E>0$), while the second integral
is referred to the reflected carriers ($-\Phi<E<0$). \cite{prb02} 
The contribution from the peak electrons is obtained as
\begin{multline} \label{den_delta}
N_\delta(\Phi) = a n_0 \, \nu(E_\delta + \Phi)\, \\ \times [ \theta(E_\delta) 
+ 2 \theta_{-\chi} \theta(-E_\delta) \,\theta(E_\delta + \Phi) ],
\end{multline}
where $E_\delta \equiv \varepsilon_\delta - \Phi_L$ is the peak location 
with respect to the barrier top. 
In Eq.~(\ref{den_delta}), the term with $\theta(E_\delta)$ gives 
the contribution at biases when the peak energy is above the barrier 
($E_\delta>0$), while the term with $\theta(-E_\delta)$ contributes 
in the opposite case $E_\delta<0$ (in the region $\chi<0$ and for 
$\Phi>-E_\delta$ only).

The electron density given by Eqs.~(\ref{den}) and (\ref{den_delta}) can now 
be substituted into the Poisson equation $d^2\Phi/dx^2=(q^2/\epsilon)N(\Phi)$ 
to find the self-consistent potential barrier position. We obtain
\begin{equation} \label{lambda}
q \ell \sqrt{ \frac{2}{\epsilon}} = \int_0^{\Phi_L} \frac{d\Phi}
{h_-^{1/2}(\Phi)} + \int_0^{\Phi_R} \frac{d\Phi}{h_+^{1/2}(\Phi)},
\end{equation}
where $h(\Phi) = \int_0^{\Phi} N d\tilde{\Phi}$, $h_-\equiv h(\chi<0)$, and 
$h_+ \equiv h(\chi>0)$. 
Integrating Eqs.~(\ref{den}) and (\ref{den_delta}) over $\Phi$, we obtain
\begin{align} \label{h}
&h = h_\delta + h_{bg}, \\
&h_\delta(\Phi) = \frac{a n_0 m}{2\pi\hbar} \{ \theta(E_\delta) \,
[v(E_\delta+\Phi) - v(E_\delta)] 
\nonumber \\ & \phantom{h_\delta(\Phi) =} \label{h_delta}
+ 2 \theta_{-\chi} \, \theta(-E_\delta) \theta(E_\delta+\Phi) \, 
v(E_\delta+\Phi) \}, 
\end{align}
\begin{widetext}
\begin{multline} \label{h_bg}
h_{bg}(\Phi) = \frac{m}{2\pi \hbar} \left\{ \int_0^\infty
[ n_{FD}(E + \Phi_L) + n_{FD}(E + \Phi_R)] \, [ v(E + \Phi) - v(E)] \,dE 
\right. \\ \left.
+  2 \int_{-\Phi}^0 
[ \theta_{-\chi} \, n_{FD}(E + \Phi_L) + \theta_{\chi} n_{FD}(E + \Phi_R)] 
\, v(E + \Phi) \, dE \right\}.
\end{multline}
\end{widetext}
Equation (\ref{lambda}) relates three important parameters: the self-consistent
barrier height $U_b$, the applied bias $U$, and the length of the conductor
$\ell$. Given any two of them, the third one can be calculated from 
Eqs.~(\ref{lambda})--(\ref{h_bg}) by making use of the boundary conditions 
(\ref{occup_L})--(\ref{occup_FD}).

\section{Current and noise}

The mean ballistic current is found as an integral over the occupation numbers
for the transmitted ($E>0$) carriers from both leads.
It can also be decomposed into two terms, $I = I_{bg} + I_\delta$, where
for each contribution we find
\begin{align}\label{curr_bg}
&I_{bg} = \frac{qA}{2\pi \hbar} \int_0^\infty 
[ n_{FD}(E + \Phi_L) - n_{FD}(E + \Phi_R) ] \,dE, \\ \label{curr_delta}
&I_\delta =  \frac{qA}{2\pi \hbar}\, a n_0 \,\theta(E_\delta).
\end{align}

Under a fixed bias condition, the fluctuations at frequencies below 
the inverse transit time can be described by a white current-noise spectrum
$S_I$. By applying the analytical method described in detail in 
Ref.~\onlinecite{prb02}, and taking into account both the background and the 
peak contributions to the noise, we obtain: 
$S_I=\tilde{S}_I^{bg} + \tilde{S}_I^\delta$, where
\begin{align}  \label{SI}
&\tilde{S}_I^{bg} = \int_{-\Phi_L}^\infty \gamma_L^2(E)\, K_{FD}(E+\Phi_L)\,dE 
\nonumber \\ &\phantom{\tilde{S}_I^{bg} =}
+ \int_{-\Phi_R}^\infty \gamma_R^2(E) \, K_{FD}(E + \Phi_R) \, dE, \\
&\tilde{S}_I^\delta = a K_0 k_BT\,\gamma_L^2(E_\delta). \label{SI_d}
\end{align}
Here the noise sources $K_{FD}(E)$ and $aK_0$ are determined by the boundary 
conditions at the leads [Eq.~(\ref{K_FD})], and the functions 
$\gamma_{L,R}(E)$ (energy-resolved shot-noise suppression factors 
\cite{prb00b}) are obtained as 
\begin{eqnarray*} \label{gamma1}
&&\gamma_L(E) = \left\{ 
  \begin{array}{l} 
  -2C_{\Delta} Q_{Lr}(E), \qquad -\Phi_L < E < 0 \\
  1 - C_{\Delta} \, Q_t(E), \qquad 0<E<\infty,
  \end{array}  \right. \\ \label{gamma2}
&&\gamma_R(E) = \left\{ 
  \begin{array}{l} 
  -2C_{\Delta} Q_{Rr}(E), \qquad -\Phi_R < E < 0 \\
  -1 - C_{\Delta} \, Q_t(E),  \qquad 0 < E < \infty,
  \end{array} \right.  
\end{eqnarray*}
where we have denoted 
$C_\Delta = c/\Delta$, 
$c = m \, n_\Delta / (2\pi\hbar)$,
$n_\Delta = n_{FD}(\Phi_L) - n_{FD}(\Phi_R)$,
\begin{align}
&Q_{Lr}(E)=\int_{-E}^{\Phi_L} v(E + \Phi) \, h_-^{-3/2} \, d\Phi, \\
&Q_{Rr}(E)=\int_{-E}^{\Phi_R} v(E + \Phi) \, h_+^{-3/2} \, d\Phi, \\
&Q_t(E) = \int_0^{\Phi_L} [v(E + \Phi) - v(E)] \, h_-^{-3/2} \, d\Phi 
\nonumber\\ & \phantom{Q_{Rr}(E)}
+ \int_0^{\Phi_R} [v(E + \Phi) - v(E)] \, h_+^{-3/2} \, d\Phi, \\
&\Delta = 2 \, [h_-^{-1/2}(\Phi_L) + h_+^{-1/2}(\Phi_R)]
\nonumber\\ & \phantom{Q_{Rr}(E)}
+ \int_0^{\Phi_L} [H_{bg}^- + H_{\delta}^-] \, h_-^{-3/2} \, d\Phi
\nonumber\\ & \phantom{Q_{Rr}(E)}
+ \int_0^{\Phi_R} [H_{bg}^+ + H_{\delta}^+] \, h_+^{-3/2} \, d\Phi,
\end{align}
where $H^+ \equiv H(\chi > 0)$, $H^-\equiv H(\chi < 0)$, and
\begin{align*} 
&H_{bg} (\Phi) = N_{bg}(\Phi)-N_{bg}(0)+ [\theta_{\chi}-\theta_{-\chi}]\, 
c v(\Phi), \\
&H_\delta (\Phi) = N_{\delta}(\Phi)- N_{\delta}(0).
\end{align*}
%
As can be easily verified for the noise power given by Eq.~(\ref{SI}), 
there is an entanglement between the background and the $\delta$ peak 
contributions (this is the meaning of a tilde we have introduced in the 
notations). The entanglement appears due to the functions $\gamma(E)$, 
in particular, due to their dependence on the steady-state functions 
$h(\Phi)$, which are the sum of both the FD and the $\delta$ peak 
contributions [see Eqs.~(\ref{h})--(\ref{h_bg})].
This is the principal difference with the time-averaged quantities, such as
the current [Eqs.~(\ref{curr_bg}) and (\ref{curr_delta})] or the electron
density [Eqs.~(\ref{den}) and (\ref{den_delta})], for which the total
value is a sum of two contributions without entanglement.
We would like to emphasize that for the current noise without Coulomb
interactions, the functions $\gamma(E)$ are constants (step functions), 
\cite{prb02} and hence the total noise is again an additive quantity 
as for the time-averaged values.

The important question is: To what extent do the cross-correlations between 
the peak and the background electrons affect the current noise?
As will be seen in the next section, the cross-correlation contribution
is of major importance at biases when the $\delta$ peak is in the vicinity of
the potential barrier top. 
Precisely this nonlinearity leads to the possibility of identifying the 
details in the energy profile of electrons from the shot-noise measurements. 

\section{Results}

For quantitative estimations and in order to illustrate the implementation 
of the results, consider the GaAs ballistic sample at $T$=4 K. 
With the assumption that the contact doping 
$1.6\times 10^{16}\,\text{cm}^{-3}$, the reduced Fermi energy $\xi \approx 10$,
and the contact electrons are degenerate, the Debye screening length 
associated with the contact electron concentration is $L_D\approx$14 nm. 
\cite{prb02}
The calculations have been carried out for the lengths of the ballistic gap
$\ell$=0.1 and 0.5$\,\mu$m, implying $\ell\gg L_D$, which is necessary
to expect the importance of the space-charge and Coulomb correlation effects.

\begin{figure}[b]
\epsfxsize=8.0cm
\epsfbox{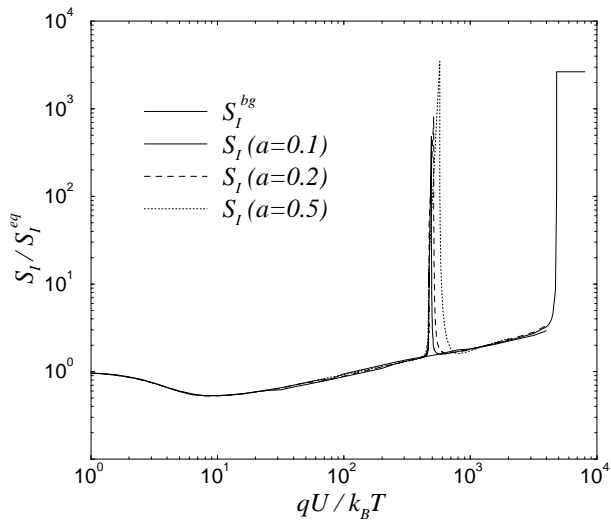}
\caption{Current-noise power $S_I$ vs applied bias $U$ for the combined
($\delta$ peak and background Fermi-Dirac) injection.
The results are shown for the $\delta$ peak at $\varepsilon_\delta$=$9k_BT$
and different amplitudes $a$. $S_I$ is normalized to the equilibrium value
$S_I^{eq}$ of the background noise. The ballistic length $\ell$=2$\,\mu$m.}
\label{peak9} 
\end{figure}

\begin{figure}
\epsfxsize=8.0cm
\epsfbox{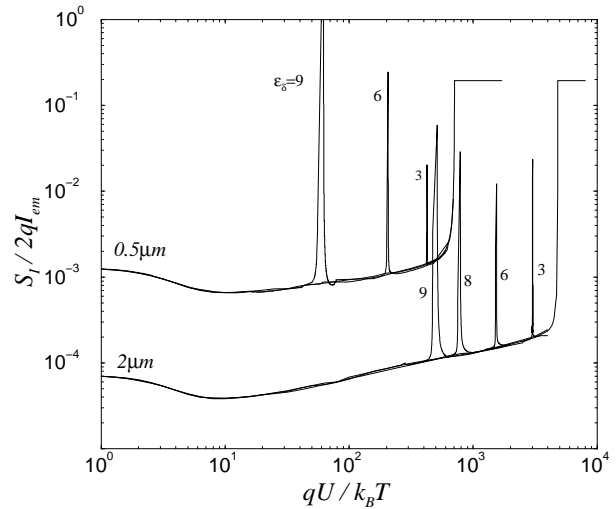}
\caption{Current-noise power $S_I$ vs applied bias $U$ for the peak of 
amplitude $a$=0.2 at different energies $\varepsilon_\delta$ 
(indicated by numbers). 
$S_I$ is normalized to the full shot noise $2qI_{em}$ in the saturation
regime, where $I_{em}$ is the saturation (emission) current from the emitter.
The results are plotted for two ballistic lengths $\ell$=0.5 and 2$\,\mu$m.}
\label{peak} 
\end{figure}

\begin{figure}
\epsfxsize=8.0cm
\epsfbox{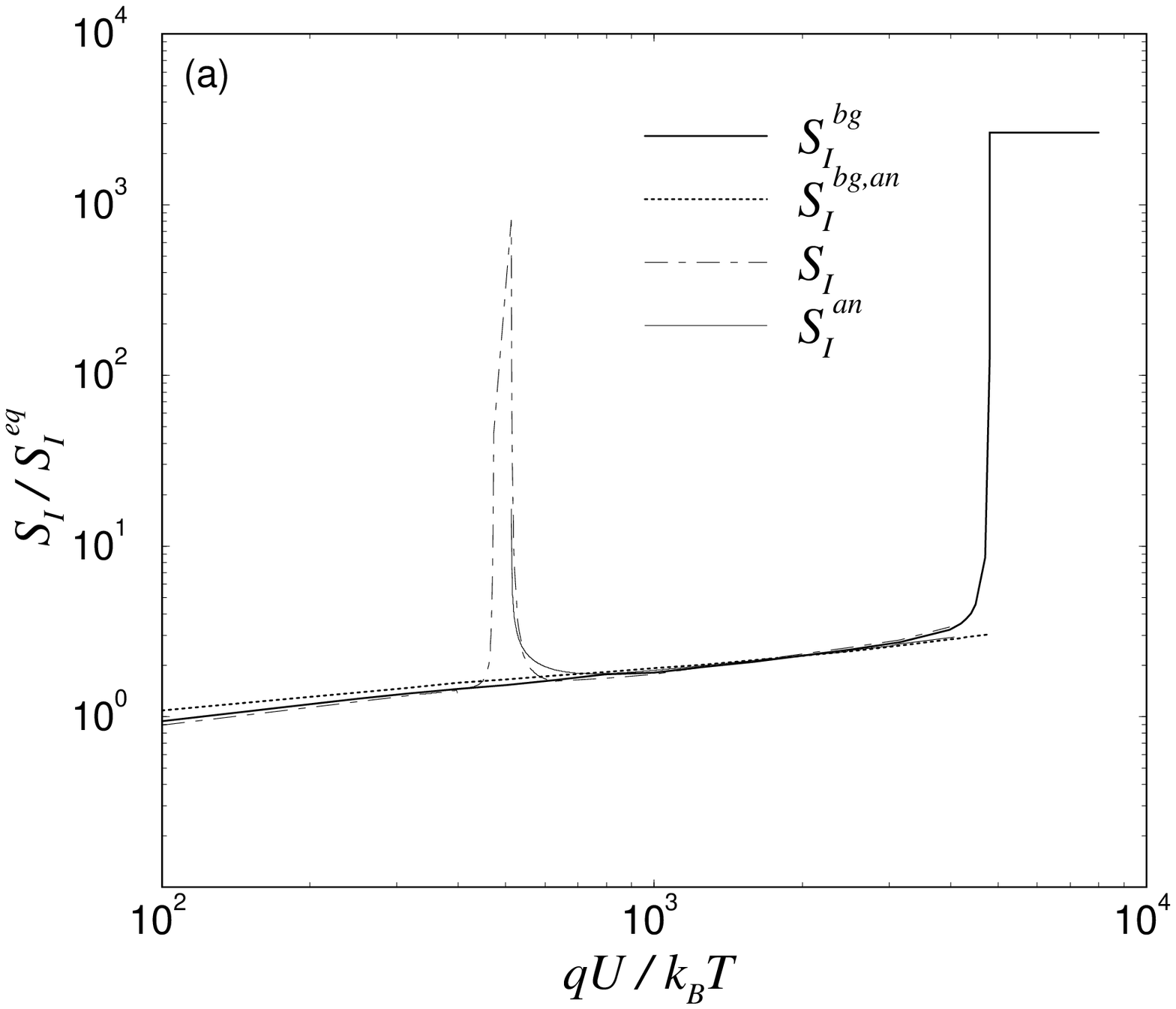}
\epsfxsize=8.0cm
\epsfbox{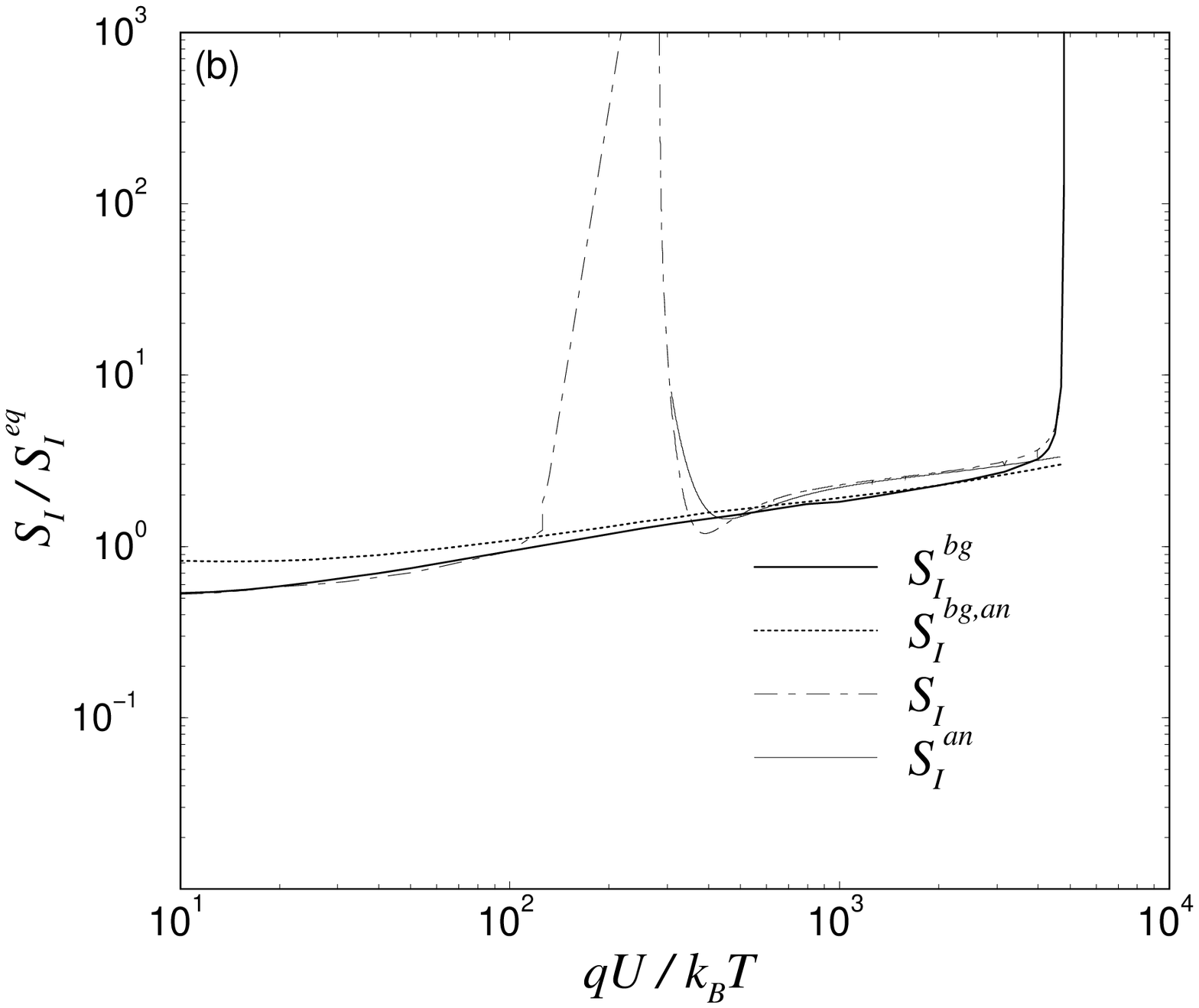}
\caption{Current noise power $S_I$ and the background contribution $S_I^{bg}$
compared with their asymptotic high-bias values $S_{I}^{an}$ and $S_I^{bg,an}$
calculated from Eq.~(\ref{asimpt}). 
All the curves are normalized to the equilibrium value $S_I^{eq}$ of the 
background noise. The parameters of the $\delta$ peak are
$e_\delta$=9, $a$=0.2 (a); $e_\delta$=11, $a$=0.5 (b).
The ballistic length $\ell$=2$\,\mu$m.} 
\label{peak11} 
\end{figure}

The calculation of the current noise power $S_I$ given by 
Eqs.~(\ref{SI}) and (\ref{SI_d}) requires the knowledge of the steady-state 
potential barrier height $\Phi_L$ and the stationary profiles $h(\Phi)$. 
To this end we solve by numerical iterations Eq.~(\ref{lambda}) for each
given bias $U$ and length $\ell$. Then we compute the noise. 
The results are presented in Figs.~\ref{peak9} -- \ref{peak11}.

First, our aim is to understand the effect of the additional peak 
electrons on the current-noise spectral density. 
Figure \ref{peak9} shows the results for the case of a $\delta$ peak fixed
at the energy $\varepsilon_\delta$=$9k_BT$ and having different amplitudes $a$.
It is seen that at low biases, when the barrier is high ($\Phi_L\approx 16$),
the peak electrons do not influence the noise since they are all reflected 
back to the emitter ($\varepsilon_\delta<\Phi_L)$. 
In this regime, $S_I\approx S_I^{bg}$.
With increasing bias $U$, the potential barrier decreases.
Then there will be a point where the energy of the peak meets the potential
barrier top. Above this point the peak electrons are no longer reflected back, 
all of them now pass over the barrier and contribute to the current 
and noise at the collector. 
But their effect on the mean current and noise is drastically different. 
If the extra current provided by the peak electrons is small with respect
to the background current (as in our example), the current-voltage 
characteristics do not change essentially.  
In contrast, the noise power changes significantly: at the bias when
$\Phi_L\approx\varepsilon_\delta$, a sharp spike in $S_I$ is observed 
(Fig.~\ref{peak9}). 
This spike is not due to the noise of the peak electrons, as one would think.
It originates from the fluctuations of the barrier top, which are induced 
by the peak electrons. This influence is very pronounced precisely 
when the condition $\Phi_L\approx\varepsilon_\delta$ holds.
Although the fraction of the peak electrons in the total current is small,
they dominate (within a narrow interval of biases at 
$\Phi_L\approx\varepsilon_\delta$) in the contribution to the electron 
density $N_b=N(\Phi=0)$ at the barrier position,
thereby changing the strength of the barrier top fluctuations.
The latter follows from the fact that the largest contribution to the electron
density $N_b$ at the barrier top comes from electrons that
virtually stop there (have zero velocity) and spend more time around 
$x$=$x_b$. 
These are the electrons from a narrow energy interval around 
$\varepsilon=\Phi_L$. 
At higher biases, $\varepsilon_\delta>\Phi_L$, the $\delta$ peak 
is shifted above the barrier towards higher energies, and the main 
contribution to $N_b$ comes again from the background electrons. Thus 
the noise power becomes $S_I\approx S_I^{bg}$ as seen from Fig.~\ref{peak9}.
So, what we really observe in Fig.~\ref{peak9} is the spike of the potential
barrier noise, which is sharply increased when the $\delta$ peak coincides
with the top of the potential barrier. In this sense the space-charge 
potential barrier helps to visualize the details in the injection currents
by means of the noise.
Note that if the potential barrier is frozen at its time-averaged value, i.e., 
when the long-range Coulomb correlations are ignored,
we obtain from the calculations that the noise spectral density is almost the
same in both cases: with and without the $\delta$ peak injection.
This indicates the importance of the long-range Coulomb correlations
in ``shot-noise spectroscopy.''

Another important feature one can observe in Fig.~\ref{peak9} is how 
the noise spike is modified with increasing density of the $\delta$ 
peak electrons. 
It is seen that the left side of the spike does not change significantly 
with $a$, while the right side is shifted towards higher biases when
$a$ increases.
This can be explained by the fact that for higher intensities of the peak,
one should apply higher biases to ``open up'' more background electrons 
in order to exceed the contribution from the peak electrons 
to the barrier top fluctuations.

An interesting question is why the delta peak injection increases
noise and not suppresses it. At the bias when the noise spike appears, 
the delta peak electrons almost stop at the barrier top and, therefore,
effectively contribute to the negative charge at the barrier top location.
This additional negative charge increases the reflection of the incoming 
electrons, that means the increase of the barrier heght.
The current fluctuation produced by the barrier increase is of the opposite 
sign in respect to the injection fluctuation. Moreover, in a small energy
range, the compensation fluctuation is much larger in absolute value with 
respect to the injection one (overcompensation effect). 
This is seen in the fact that the function $\gamma(E)\to -\infty$ 
as the energy approaches the barrier top $E\to 0$. \cite{prb00,cm99} 
Since the noise is calculated as a square of fluctuation, the result is 
the increase of noise, although it is due to the suppression effect of current 
fluctuations. 

Figure \ref{peak} illustrates the shot-noise spectroscopy effect. 
It is seen that when the energy of the peak electrons changes, the noise spike
shifts along the background noise curve, since the bias at which the condition
$\varepsilon_\delta\approx\Phi_L$ is met changes. 
This allows one, in principle, to identify the peak energy if one knows
the correspondence between the bias and the barrier height for the background
noise.

Finally, we have also verified the validity of our numerical algorithm 
by comparing $S_I$ calculated from the full set of Eqs.~(\ref{SI}) and 
(\ref{SI_d}) with the analytical results obtained in the asymptotic 
high-bias limit: \cite{pe02}
\begin{equation}  \label{asimpt}
S_I^{an} = 2q I_{bg} \frac{9k_B T}{qU} \left[ 1 - \sqrt{\pi} w g_1 + w^2 g_2 + 
\tilde{a}(\sqrt{e_\delta} - w)^2\right],
\end{equation}
where $w = (\sqrt{\pi} g_1 + \tilde{a}/\sqrt{e_\delta})/(2 g_2)$, 
$\tilde{a} = a / {\cal F}_1(\alpha)$ is the ratio between the current of 
the peak and the current of the background, 
$g_1 = {\cal F}_{1/2}(\alpha) / {\cal F}_1(\alpha)$ and 
$g_2 ={\cal F}_0(\alpha) / {\cal F}_1(\alpha)$ are the constants dependent on 
the degeneracy of the injected electrons ($g_1$=$g_2$=1 for nondegenerate 
electrons), ${\cal F}_j$ are the Fermi-Dirac integrals of index $j$, and 
$\alpha = \xi - \Phi_L / k_B T$ and 
$e_\delta = E_\delta/ k_B T$ are the dimensionless positions of 
the Fermi energy and the $\delta$ peak with respect to the potential barrier, 
respectively. 

The results of the comparison are in good agreement, as Figs.~\ref{peak11}(a) 
and \ref{peak11}(b) shows. For moderate peak amplitudes, 
the exact solutions follow closely the asymptotic curves to the right of 
the noise spike. Moreover, for this case the results for $qU\gg E_\delta$ 
are almost the same in both cases: with and without the $\delta$ peak 
[Fig.~\ref{peak11}(a)]. 
For high peak amplitudes [Fig.~\ref{peak11}(b)], the asymptotic curves
differ by a small value corresponding to the noise contribution from the peak 
electrons, non-negligible in this case.

It is remarkable, that the asymptotic theory describes quite well not only 
the region $qU\gg E_\delta\gg\Phi_L$, where the $\delta$ peak is much higher 
than the barrier, but also the right side of the spike at 
$E_\delta\sim\Phi_L$, 
where the peak is close to the barrier position (slightly above it). 
The left side of the peak is, however, beyond the asymptotic theory, since
in this bias range the $\delta$ peak electrons are reflected from the barrier,
and no asymptotic theory exists for this case.

Another important feature is that for the case of a high amplitude peak,
there exists a small range of biases for which $S_I<S_{bg}$ 
[Fig.~\ref{peak11}(b)]. This means that the current noise may be reduced 
by an additional injection (additional noise).
We relate this noise suppression phenomenon to the existence of a specific 
``noiseless'' energy $E^*$ lying above the barrier \cite{prb02} for which 
$\gamma(E^*)\approx 0$. 
At the bias when $E_\delta$ crosses $E^*$, the noise of the $\delta$ peak 
vanishes, $\tilde{S}_I^\delta\approx 0$, while the background noise is reduced
by the peak electrons by virtue of the barrier fluctuations.

\section{Conclusions}

We have investigated the shot noise in a space-charge limited ballistic 
conductor under the condition of an additional (to the thermal equilibrium) 
$\delta$ peak injection from one of the contacts.
Coulomb interactions are shown to act as an energy analyzer of the profile 
of injected electrons, and help to visualize the energy peak in the injection 
current by means of the shot noise.
The injection peak is not seen in the time-averaged measurements or 
shot-noise measurements under the conditions when interactions are ineffective.

By measuring the current-noise spectral density as a function of bias,
one can observe a sharp spike in the noise at a certain bias, at which
the energy peak coincides with the potential barrier top.
This spike is a signature of electron-electron interactions.
It gives a direct link between the peak and the barrier positions,
and can reveal one position when the other is known.
For instance, if one knows the injection energy $E_{\delta}$ of the $\delta$ 
peak electrons, one can obtain the bias at which the barrier height 
$\Phi_b\approx E_{\delta}$, thereby revealing the space-charge-limited
conduction and the amount of the space charge.
Vice versa, if one knows the space-charge parameters and the barrier 
position as a function of bias, one can analyze the energy-resolved injection
current.

Thus, Coulomb interactions in ballistic structures 
are of interest from several points of view:
On one hand, they lead to the shot-noise suppression
that may be important for applications.
On the other hand, they offer the possibility of using the shot-noise
measurements as a tool to deduce
important information about the properties of nonequilibrium carriers
in nanoscale structures with hot-electron emitters,
resonant-tunneling-diode emitters, superlattice emitters, etc.,
not otherwise available from time-averaged measurements.
The validity of our theory can be tested experimentally in currently
accessible semiconductor ballistic structures, in which the current
is limited by the space charge.

\begin{acknowledgements}
This work was partially supported by the DGICYT and FEDER under 
Grant No.\ BFM2002-01267.
O.M.B. acknowledges support by the Ministerio de Ciencia y
Tecnolog\'{\i}a of Spain through the ``Ram\'on y Cajal'' program.

\end{acknowledgements}

\end{document}